\documentclass[11pt,twoside]{article}

\usepackage{asp2006}
\usepackage{epsf}
\usepackage{psfig}
\usepackage{lscape}

\begin{document}
\title{The Effect of Dry Mergers on the Color-Magnitude Relation}
\author{Rosalind E. Skelton,$^1$ Eric F. Bell,$^1$ Rachel S. Somerville$^{1,2}$}
\affil{$^1$Max-Planck-Institut f\"ur Astronomie,
K\"onigstuhl 17, D-69117 Heidelberg, Germany\\$^2$Space Telescope Science Institute, 3700 San Martin Drive, Baltimore, MD 21218}

\begin{abstract} 
We investigate the effect of gas-poor (so-called ``dry'') mergers on the color-magnitude relation (CMR) of early-type galaxies through a simple toy model and compare with low-$z$ observations from the Sloan Digital Sky Survey (SDSS). The observed red sequence shows a tilt towards bluer colors and a decrease in scatter at the bright end. These characteristics are predicted by our model, based on merger trees from a semi-analytic model of galaxy formation. We assume galaxies move onto a ``creation red sequence'' when they undergo major gas-rich mergers. Subsequent dry mergers move galaxies along the relation by increasing their mass, but also make them slightly bluer. This occurs because bright galaxies are most likely to merge with one of the more numerous fainter and consequently bluer galaxies that lie further down the relation. Bright galaxies undergo a higher fraction of dry mergers than faint galaxies, which causes a change in the slope of the CMR. A more realistic model that includes scatter in the initial relation shows that dry merging causes a tightening of the CMR towards the bright end. The small scatter in the observed CMR thus cannot be used to argue against significant mass growth from dry merging.
\end{abstract}


\section{Introduction}
Galaxies can be separated into two main classes, occupying different regions in color-magnitude space \citep{RS_Strateva01,RS_Blanton03}. The gas-rich, star-forming galaxies form a broad distribution known as the blue cloud. The early-type or spheroidal galaxies are gas-poor and have low levels of star formation. These galaxies lie along a tight color-magnitude (or mass) relation known as the red sequence, which is driven to first order by a metallicity-mass relation \citep{RS_Faber73, RS_Kodama97, RS_Gallazzi06}. The amount of mass in red galaxies has approximately doubled since $z\sim1$ \citep[e.g][]{RS_Bell04,RS_Faber07} whereas the mass in the blue cloud remains roughly constant. Galaxy merging is thought to play an important role in moving galaxies from the blue cloud onto the red sequence, through morphological transformation \citep[e.g.][]{RS_Toomre77,RS_Barnes96} and the quenching of star formation. The most massive ellipticals are more likely to build up from mergers between galaxies which already lie on the CMR, however. Such mergers are observed in the local Universe \citep[e.g.][]{RS_vanDokkum05,RS_Bell06a,RS_McIntosh08} but the resulting mass growth is difficult to constrain. The rate at which they occur is uncertain, to a large extent due to the uncertainty in merging time-scale, and indirect means of measuring it, through the evolution of the mass-size relation \citep{RS_vanderWel08} or number density \citep[e.g.][and references therein]{RS_Faber07,RS_Cool08} for example, are thus far are inconclusive.
\par
The CMR provides a further avenue to explore, with both the slope and scatter giving insight into the formation history of elliptical galaxies. The relation is generally assumed to be linear, though there is some evidence for a change in slope with magnitude \citep[e.g.][]{RS_Baldry04,RS_Ferrarese06}. Previous work on the scatter in the CMR suggested that ellipticals formed at high redshifts, evolving passively thereafter \citep[][BKT98 hereafter]{RS_BKT98} though recent models allow for the continuous build-up of the red sequence through the quenching of blue cloud galaxies \citep{RS_Harker06, RS_Ruhland09}. BKT98 argued that the amount of mass growth due to dry merging was limited to a factor of 2--3 because such mergers would cause a flattening of the relation and an increase in scatter, contradicting the small scatter measured in clusters such as Coma. In this contribution we show that the CMR for local field galaxies from the SDSS has a change in slope (\S2). We attribute this effect to dry mergers, using a simple toy model to investigate how they influence the red sequence (\S3). Further details can be found in Skelton, Bell, \& Somerville (submitted to ApJ).

\section{Observations}
We select a subsample of galaxies from the SDSS Data Release 6 \citep[DR6,][]{RS_Adelman-McCarthy08} using the New York University Value-Added Galaxy Catalog \citep[NYU-VAGC;][]{RS_Blanton05b}. We choose galaxies in a thin redshift slice ($0.0375 < z < 0.0625$) with Galactic extinction corrected \citep{RS_Schlegel98} Petrosian magnitudes $m_r < 17.77$ (72646 objects). This range in redshift provides a significant number of bright galaxies but is narrow enough to avoid the need for volume and evolution corrections. We use the S\'ersic magnitudes provided in the NYU-VAGC as an estimate of total magnitude and the SDSS model magnitudes, determined with an equivalent aperture in all bands, as the most reliable estimator of color. We $k$-correct to rest-frame $z = 0.1$ bandpasses using \texttt{kcorrect\_v4\_1} \citep{RS_BlaRow07}. To isolate the red sequence we apply a cut in concentration of $C \ge 2.6$, where $C = R_{90}/R_{50}$ with $R_{90}$ and $R_{50}$ the radii enclosing 90\% and 50\% of the Petrosian flux, respectively. The remaining sample contains 29017 galaxies and is complete for $M_r \la -18.3$~mag. 
\par
\begin{figure}[!ht]
\plotone{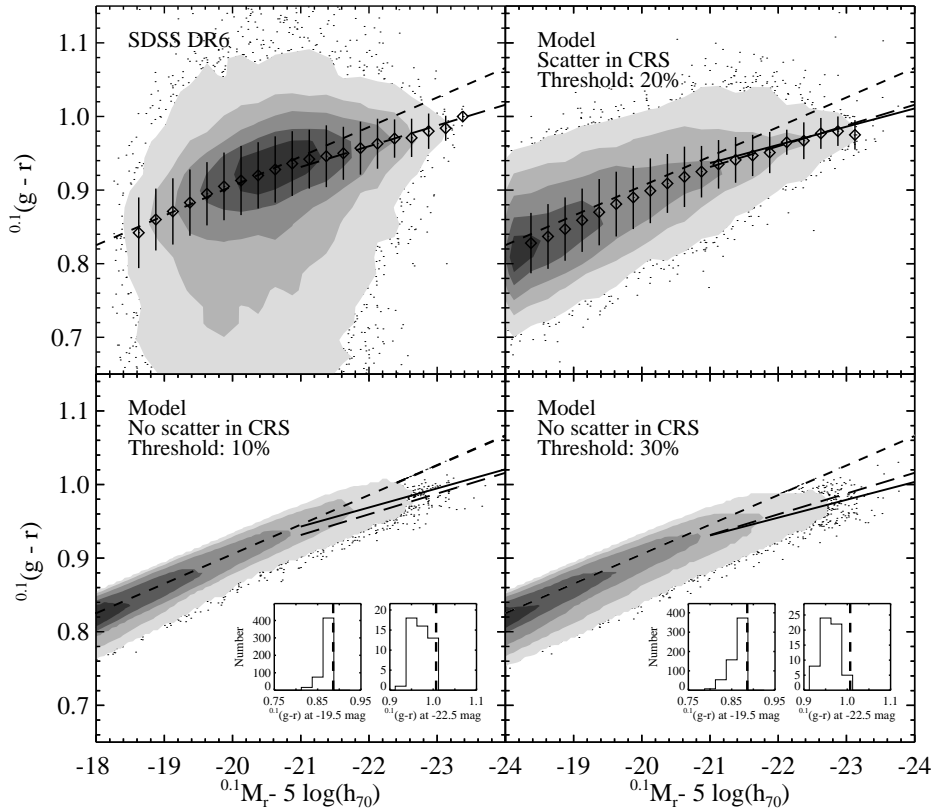}
\caption{The upper left panel shows the CMD of galaxies with concentrations of $C \geq 2.6$ from the SDSS DR6. The upper right panel shows the red sequence for a model which includes scatter in the CRS, with a gas fraction threshold of 20\%. The mean and scatter of Gaussian fits are shown as diamonds and bars. The lower panels show models with gas fraction thresholds of 10 and 30\% with no scatter in the initial relation. In all panels the contours enclose 2, 10, 25, 50 and 75\% of the maximum value. The short dashed lines show the fit to the observed means for magnitudes fainter than $M_r = -21$~mag, extended over the whole magnitude range to illustrate the change in slope at the bright end. Long dashed and solid lines show the fit to the bright end ($M_r < -21$~mag) of the observed and model relations respectively. The inset histograms show the distribution of colors in two magnitude bins 0.1~mag wide, centered on $^{0.1}M_r=-19.5$ and $^{0.1}M_r=-22.5$~mag. \label{cmd.fig}} 
\end{figure}

The CMD for these galaxies (upper left panel of Fig.~\ref{cmd.fig}) shows a change in slope with magnitude, flattening at the bright end. We fit a Gaussian function to the distribution of colors in each magnitude bin of 0.25~mag and fit straight lines to the means above and below $M_r = -21$~mag. The faint-end fit given by $^{0.1}(g - r) = 0.10 - 0.04 ^{0.1}M_r$ and average scatter (0.046~mag) are used as the input creation red sequence for the model, as described below. 

\section{A Toy Model for Dry Merging}
We use a simple toy model to isolate the effect of dry merging on the colors of galaxies on the red sequence, taking a similar approach to BKT98. We differ in that we use galaxy merger histories from an up-to-date model of galaxy formation for all haloes with ${\rm{log}}_{10}[{\rm M_{halo}}/{\rm M_{\sun}}] > 11.7$, rather than just cluster-sized haloes. Furthermore, we suppose that major wet mergers quench star formation and move galaxies onto a ``creation red sequence'' (CRS), whereas the BKT98 model assumes the red sequence is in place at some formation epoch. Galaxy merger trees and the masses and gas fractions of galaxies are extracted from the semi-analytic galaxy formation model (SAM) of \citet[][]{RS_Somerville08} embedded in the $\Lambda$CDM hierarchical framework. 
\par
We consider major mergers with mass ratios\footnote{The mass used is the total baryonic mass plus the dark matter mass within twice the characteristic NFW scale radius (see Somerville et al.~2008 for details).} between 1:1 and 1:4 to be sufficient at quenching star formation if either of the progenitor galaxies has a cold gas fraction\footnote{Cold gas fraction is defined as the ratio of cold gas to total baryonic mass.} above some threshold. In order to compare directly with observations we have used the measured faint-end slope and zeropoint of the observed red sequence to specify the remnant's color ($^{0.1}(g - r) = 0.10 - 0.04 ^{0.1}M_r$, short dashed lines in Fig.~\ref{cmd.fig}). The magnitude of the remnant galaxy is found from the total stellar mass of the two progenitors using the M/L ratio of low-$z$ red sequence galaxies produced by the SAM ($^{0.1}M_r = 2.87 - 2.22 {\rm{log}_{10}[M_*}/{\rm M_{\sun}]}$). Subsequent dry merging produces remnant galaxies with colors and magnitudes determined by the simple combination of the progenitor colors and magnitudes. A more realistic model includes scatter in the CRS, with the width of the relation given by the average observed scatter for faint galaxies (0.046~mag). 
\par
In Fig.~\ref{cmd.fig} we compare the observed red sequence (upper left panel) with the CMR produced by the model, using gas fraction thresholds ranging from 10 to 30\%. A linear regression to the bright end ($M_r < -21$~mag) of the model distribution is shown as a solid line in each case. There is a clear tilt towards a shallower slope for bright galaxies, with the slope and break point sensitive to the chosen gas fraction threshold. The observed bright-end slope is bracketed by models with thresholds of 10\% and 30\% (lower panels). The inset histograms show that the distribution peaks on the CRS (dashed line) in the faint bins because most of these galaxies have not had dry mergers. In contrast, most bright galaxies have undergone dry mergers and are predicted to lie significantly bluewards of the CRS at the present day. A lower gas fraction threshold results in more wet merger remnants lying directly on the CRS, thus the slope at the bright end changes less dramatically and the break occurs at brighter magnitudes. The upper right panel shows the resulting red sequence for the model with scatter in the CRS and a gas fraction threshold of 20\%. We fit a Gaussian to the color distribution in each magnitude bin of 0.25~mag, as for the observations. The slope at the bright end decreases in the same way as for the models without scatter, while the relation becomes tighter towards the bright end. The small fraction of dry mergers taking place at all magnitudes result in a small zeropoint offset. 

\section{Conclusions}
The existence of a tight CMR over a wide range in magnitude has been used to argue against the importance of dry mergers because they were expected to increase the scatter and flatten the relation (BKT98). We show that the CMR for local field galaxies from the SDSS has a change in slope at the bright end and that a toy model for dry merging in an hierarchical Universe produces a red sequence that is consistent with these observations. In models without scatter dry mergers shift galaxies at the bright end towards bluer colors and increase the width of the relation. There is little change at the faint end because most mergers occurring there are wet. This results in a change in slope rather than a flattening of the whole relation, predicted by BKT98. They assumed that the red sequence formed at a given time and that subsequent merging at all stellar masses was dry, moving the entire population bluewards, whereas we associate the build-up of the red sequence with merging events. The change in slope and magnitude at which the break occurs depend strongly on the assumption of a gas fraction threshold below which mergers are assumed to be dry. Thresholds of 10\% and 30\% bracket the observed relation. Including scatter in the CRS, we find a reduction in scatter at the bright end as a consequence of the central limit theorem. The scatter in the observed relation is slightly smaller than in the model, however we have not accounted for differences in age or metallicity. We have assumed the CRS has the same scatter as the faint end of the observed red sequence, which has a contribution from the aging of the stellar populations \citep{RS_Gallazzi06}, thus the scatter could be even smaller than predicted if dry mergers occur soon after the arrival of galaxies on the red sequence. 

\acknowledgements RES would like to thank the conference organisers for a very interesting and well-run workshop. This work makes use of the Sloan Digital Sky Survey (http://www.sdss.org/).

\end{document}